\documentclass[10pt,aps,prl,twocolumn,showpacs,showkeys,amsmath,floatfix,superscriptaddress,preprintnumbers]{revtex4-1}
\usepackage{color,graphicx}
\usepackage{mathptmx}                
\usepackage{dcolumn}                 
\usepackage{bm}                      
\usepackage{ulem,soul}
\usepackage[pdf]{pstricks}
\usepackage{nicefrac}
\usepackage{multirow}
\DeclareMathAlphabet{\mathpzc}{OT1}{pzc}{m}{it}

\begin{document}


\title{General predictions for the neutron star crustal moment of inertia}

\author{Thomas Carreau}
\author{Francesca Gulminelli}
\affiliation{CNRS, ENSICAEN, UMR6534, LPC ,F-14050 Caen cedex, France}

\author{J\'er\^ome Margueron}
\affiliation{Institut de Physique Nucl\'eaire de Lyon, CNRS/IN2P3, Universit\'e de Lyon, Universit\'e Claude Bernard Lyon 1, F-69622 Villeurbanne cedex, France}

\date{\today}

\begin{abstract}
The neutron star crustal EoS and transition point properties are computed within a unified meta-modeling approach. 
A Bayesian approach is employed including two types of filters: bulk nuclear properties are controlled from low density effective field theory (EFT) predictions as well as the present knowledge from nuclear experiments, while the surface energy is adjusted on experimental nuclear masses. Considering these constraints, a quantitative prediction of crustal properties can be reached with controlled confidence intervals and increased precision with respect to
previous calculations: {$\approx 11\%$} dispersion on the crustal width and {$\approx 27\%$} dispersion on the fractional moment of inertia. 
The crust moment of inertia is also evaluated as a function of the neutron star mass,
{ and predictions for mass and radii are given for different pulsars.} 
 The possible crustal origin of Vela pulsar glitches is discussed within the present estimations of crustal entrainment, disfavoring a large entrainment phenomenon if the Vela mass is above $1.4M_\odot$. 
Further refinement of the present predictions requires a better estimation of the high order isovector empirical parameters, e.g. $K_{sym}$ and $Q_{sym}$, and a better control of the surface properties of extremely neutron rich nuclei.
\end{abstract}

\maketitle


\begin{figure}[h]
    \centering
    \includegraphics[scale=0.4]{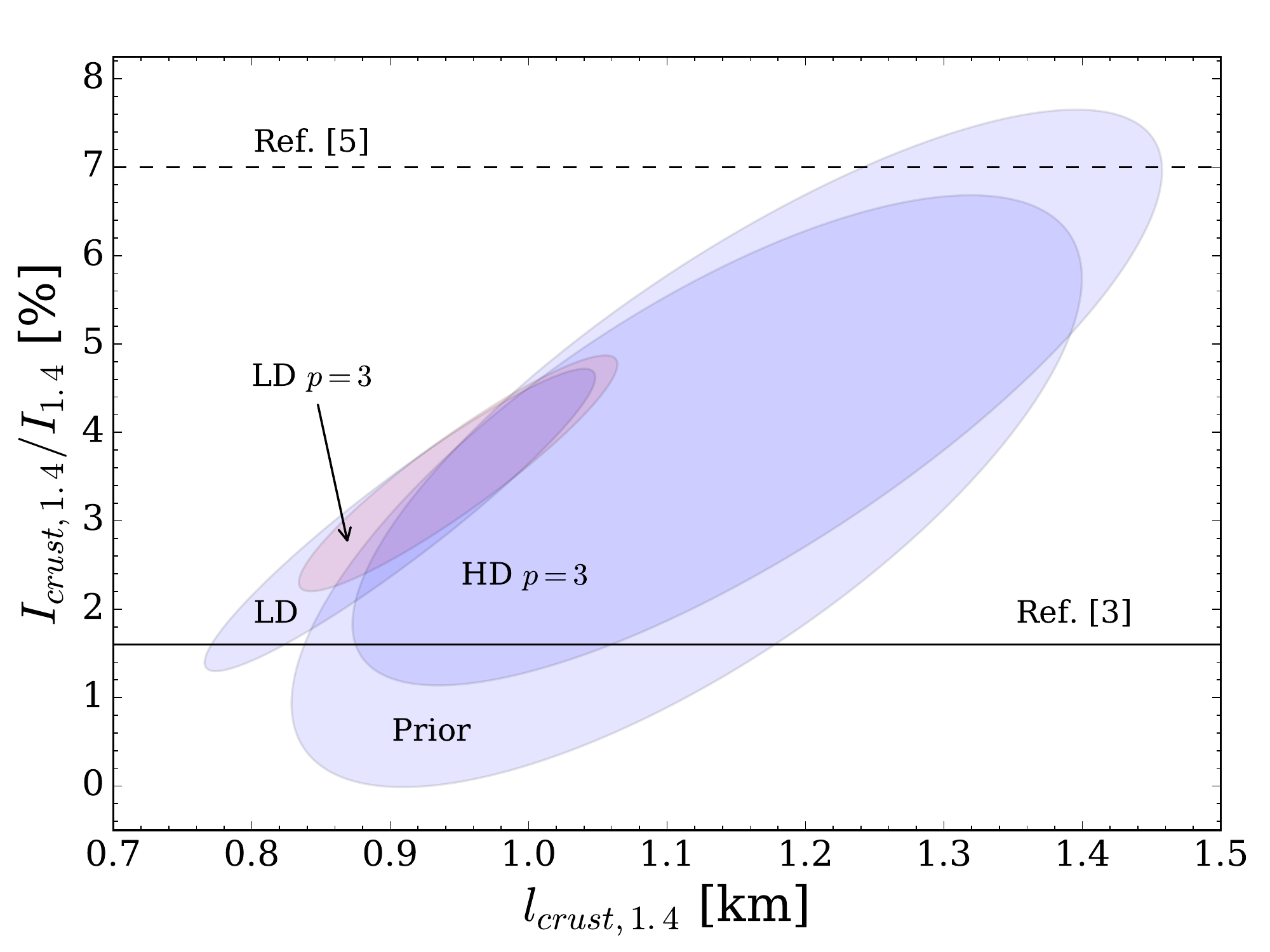}\\
     \vspace{-6pt}
    \includegraphics[scale=0.37]{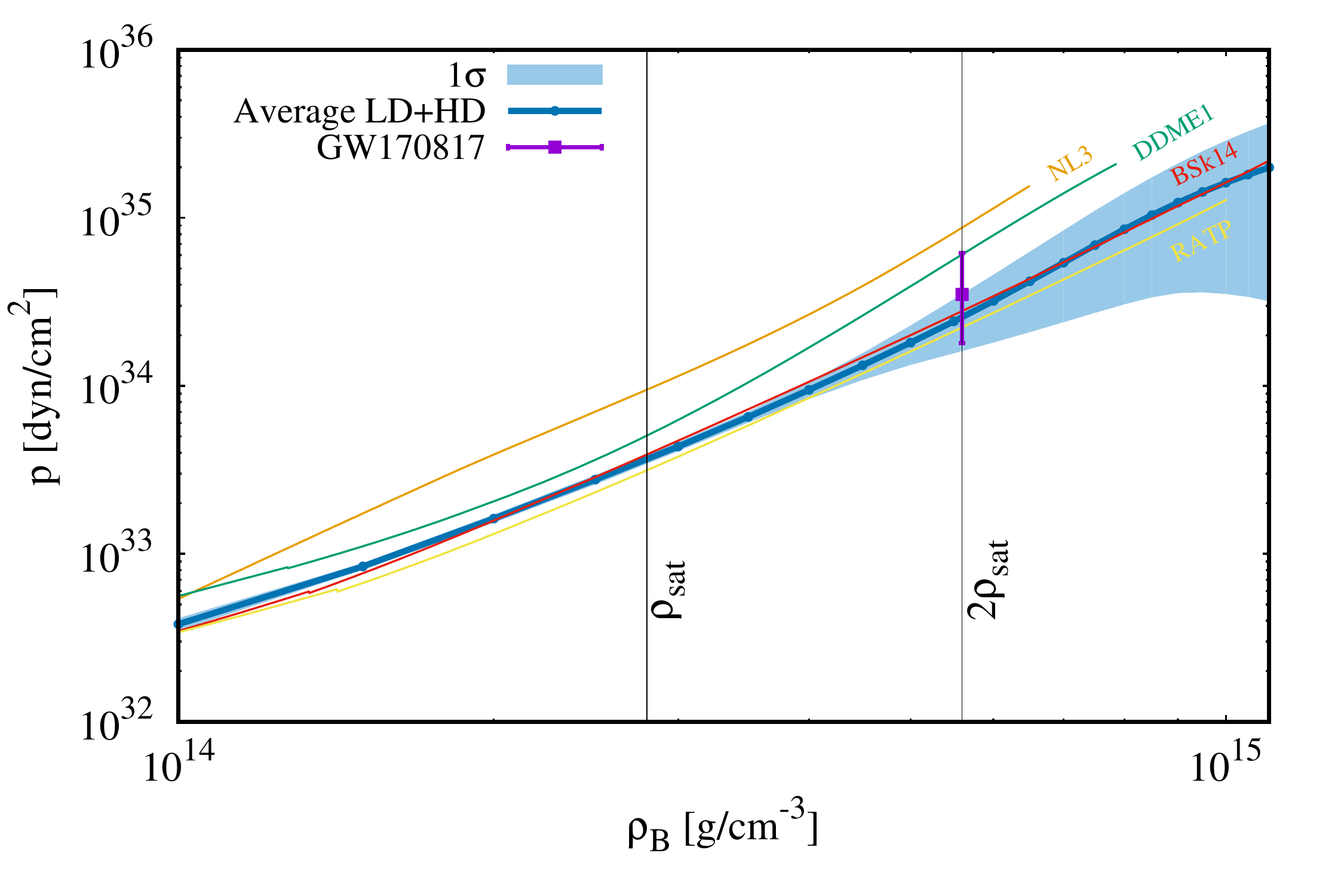}\\
     \vspace{-15pt}
\caption{(Color online)
Top: 
1$\sigma$ confidence ellipse for the crust thickness $l_{crust}$ and the fraction of crust moment of inertia $I_{crust}/I$ for a 1.4$M_\odot$ neutron star with different filters (see text). Minimal values needed to justify Vela glitches without~\cite{Link1999} and with~\cite{Andersson2012} entrainment are represented.{ Bottom: behavior of the equation of state retained by this study compared to some popular models. The recent constraint from GW170817 \cite{GW170817} is also given.}}\label{fig:sigma_ntpt}
     \vspace{-15pt}    
\end{figure}

The standard theory of pulsar glitches, this sudden  spin-up of the rotational frequency of a compact star observed in almost 200 different pulsars since their discovery
\cite{Espinoza2011}, assumes that the observed phenomenon originates from an abrupt transfer of angular momentum from the neutron superfluid to the solid crust of the star, due to the unpinning of the superfluid vortices from the crystal lattice \cite{Haskell2015}. For this mechanism to justify the large glitches observed in { some pulsars such as Vela}, the neutron star crust must be sufficiently thick to store a significant amount of angular momentum. The corresponding fraction of crust moment of inertia
$I_{crust}/I$ can be estimated \cite{Link1999,Delsate2016,Andersson2012} in a range going from 1.6 \%  up to 15 \%, depending on the importance of the effect of crustal entrainment, which is currently under debate~\cite{Martin2016,Watanabe2017}.

A reliable estimation of the crust thickness and of the associated moment of inertia is therefore crucially needed to validate the crustal origin of pulsar glitches. 
This quantity is also a key parameter for the simulations of neutron star cooling~\cite{Page2013}. 
For this estimation, constraints from low energy nuclear physics appear more promising than direct constraints from astrophysics~\cite{Newton2014,GW170817}.
 Indeed, the only poorely known parameter for the determination of the crustal thickness of a neutron star is the nuclear EoS and, most important, the density and pressure at the transition point from the solid crust to the liquid core~\cite{Piekarewicz2014}.

\begin{table*}[ht]
\centering
\label{my-label}
\begin{tabular}{lcccccccccccc}
\cline{2-5}
\hline
\hline
\multicolumn{1}{c}{} & 
\multicolumn{2}{c|}{$n_t$ (fm$^{-3}$)} & 
\multicolumn{2}{c|}{$P_t$ (MeV/fm$^3$)} &
\multicolumn{2}{c|}{$\rho_{c,1.4} (\times 10^{14})$ (g/cm$^3$)} &
\multicolumn{2}{c|}{$R_{1.4}$ (km)} &
\multicolumn{2}{c|}{$l_{crust,1.4}$ (km)} &
\multicolumn{2}{c}{$I_{crust,1.4}/I_{1.4}$ (\%)} \\ 
\multicolumn{1}{c}{} & 
\multicolumn{1}{c}{Average} & \multicolumn{1}{c|}{$\sigma$} & 
\multicolumn{1}{c}{Average} & \multicolumn{1}{c|}{$\sigma$} & 
\multicolumn{1}{c}{Average} & \multicolumn{1}{c|}{$\sigma$} & 
\multicolumn{1}{c}{Average} & \multicolumn{1}{c|}{$\sigma$} & 
\multicolumn{1}{c}{Average} & \multicolumn{1}{c|}{$\sigma$} & 
\multicolumn{1}{c}{Average} & \multicolumn{1}{c}{$\sigma$} \\
\cline{2-5} 
\hline
prior & 0.089 & 0.037 & 0.310 & 0.340 & 6.661 & 1.102 & 12.77 & 0.61 & 1.13 & 0.29 & 3.40 & 3.34 \\
HD    & 0.075 & 0.032 & 0.392 & 0.328 & 6.455 & 1.013 & 12.80 & 0.65 & 1.17 & 0.29 & 4.39 & 3.26 \\
LD    & 0.074 & 0.011 & 0.364 & 0.122 & 7.820 & 1.075 & 11.94 & 0.42 & 0.95 & 0.11 & 3.54 & 1.33 \\
LD+HD & 0.077 & 0.010 & 0.389 & 0.111 & 6.756 & 0.606 & 12.47 & 0.25 & 1.03 & 0.10 & 4.50 & 1.25 \\ 
\hline
\hline
\end{tabular}
\caption{Average value and standard deviation of the transition density $n_t$, transition pressure $P_t$, central mass density $\rho_c$, radius $R$, crust thickness $l_{crust}$, and crustal fraction of moment of inertia for a 1.4$M_\odot$ neutron star for different filters. We impose $p=3$.}
\label{tab:ntpt}
\end{table*}

In this Letter, we present a unified EoS treatment~\cite{Fortin2016,Sharma2015,Shen1998}, where the core and crust EoS are built within the same {functional.} To evaluate the uncertainties induced by the incomplete knowledge of the EoS, a 
{meta-modeling technique is used. It consists in generating a}
large set (100 millions) of 
models { with fully independent model parameters} 
using the strategy proposed in Refs.~\cite{Margueron2018a,Margueron2018b}. The  probability distribution of the 
 parameters is evaluated in a Bayesian approach, by constraining energy and pressure in low density homogeneous matter from { a many-body perturbation theory (MBPT) based on two and three-nucleon chiral EFT interactions at N$^3$LO and generating band predictions in isospin-symmetric and neutron matter~\cite{Drischler2016}.}
 The priors  are determined from nuclear phenomenology and a similar meta-modeling technique was already employed in Ref.~\cite{Steiner2013}.

The nuclear experimental and low density theoretical uncertainties can thus be translated into a confidence ellipse~\cite{Friendly2013} for the crustal thickness and moment of inertia. This is shown by the "LD" correlation curve of Fig.~\ref{fig:sigma_ntpt} which anticipates our main result.

The predictions considering only the nuclear experimental constraints included as uncorrelated parameter set are labelled as "Prior" while the predictions from EoS models which further satisfy basic physical constraints at high density -- see more details in the following -- are labelled "HD". 
Fig.~\ref{fig:sigma_ntpt} shows that the "LD" prediction is considerably less dispersed than the "Prior" or "HD" ones. {The $2\sigma$ surface of the complete (LD+HD) prediction including all constraints corresponds to a 84\% confidence level~\footnote{This value deviates from the standard 90\% because of non-gaussianity of the probability distribution.}, and the corresponding EoS is represented in the lower part of the Figure.}

As discussed above, the crust properties  require the knowledge of the crust-core transition density and pressure. 
They have been calculated by many authors using different versions of the density functional theory~\cite{Ducoin2011,Piekarewicz2014}. 
Most calculations are based on the thermodynamical spinodal, while this method  provides only a qualitative estimation of the crust-core transition~\cite{Pais2016,Ducoin2007,Xu2009}. 
Indeed, the transition occurs when the inhomogeneous phase becomes energetically favored over the homogeneous one~\cite{Baym1971}, which is governed by the interplay between the surface tension and the Coulomb energy. As a matter of fact, none of these terms contribute to the thermodynamical spinodal.  
A better estimation is obtained from the so-called dynamical spinodal~\cite{Pethick1995}, which corresponds to the instability border with respect to finite size density fluctuations.
Such calculations have however been performed for a small set of models~\cite{Ducoin2007,Xu2009,Ducoin2011,Pais2016}.

Following Ref.~\cite{Margueron2018a}, the generated meta-models are characterized by a set of empirical parameters { $\{\vec{P_\alpha}\}=\{n_{sat},K_{sat},Q_{sat},Z_{sat},E_{sym},L_{sym},Q_{sym},Z_{sym},m^*,\Delta m^*,b\}$,} corresponding to the successive density derivatives at saturation of the uniform matter binding energy in the isoscalar and isovector channels. 
{ They characterize the density dependence of the energy in symmetric matter, 
 and of the symmetry energy. 
}
An expansion up to the fourth order is necessary and sufficient to guarantee an excellent reproduction of existing functionals up to $4n_{sat}$, where $n_{sat}$ is the saturation density of nuclear matter~\cite{Margueron2018a}.
Two additional parameters rule the density dependence of the effective mass $m^*$ and the effective mass splitting $\Delta m^*$, and an extra $b$ parameter enforces the correct behavior at zero density.
{ This last parameter measures the low density deviation from a Taylor expansion at saturation, and turns out to be completely uninfluential  in this study (see Fig.~\ref{fig:cor} below).}

In 
the neutron star crust, the meta-modeling is extended with a surface term, 
validated through comparisons with Thomas-Fermi calculations~\cite{Ravenhall1983},
\begin{equation}
\sigma_s(x)=\sigma_0\frac{2^{p+1}+b_s}{x^{-p}+b_s+(1-x)^{-p}}, \label{eq:surface}
\end{equation}
where $x$ is the cluster proton fraction, see also Refs.~\cite{Lorentz1993,Newton2013}.
The crust composition is then variationally determined within the compressible liquid drop model (CLDM) approximation \cite{Baym1971,Douchin2001,Sharma2015,Shen1998}.

The expression (\ref{eq:surface}) for the surface tension requires three additional parameters. 
$\sigma_0$ and $b_s$ are adjusted to reproduce experimental masses of spherical magic and semi-magic nuclei: $^{40,48}$Ca, $^{48,58}$Ni, $^{88}$Sr, $^{90}$Zr, $^{114,132}$Sn, and $^{208}$Pb~\footnote{Enlarging the set of mass data does not modify the results.}.
The isovector surface parameter $p$ determines the behavior of the surface energy for extreme isospin values, and  it cannot be determined from experiments.
In the following, we consider two different choices: either a fixed value $p=3$, as suggested in Ref.~\cite{Lorentz1993}, or including  
$p$ in the parameter set $\{\vec{P_\alpha}\}$.

For each set of uniform matter parameters $\{\vec{P_\alpha}\}$, our fit provides optimal values for $\sigma_0$ and $b_s$ and the resulting $\chi^2$ enters the Bayesian likelihood probability defined as,
\begin{eqnarray}
p_{lik} (\{\vec{P_{\alpha}}\}) = \mathpzc{N} \, e^{-\frac 1 2 \chi^2(\{\vec{P_\alpha}\})} \, \prod_{\alpha} \; g(\{\vec{P_\alpha}\}),
\label{eq:probalikely}
\end{eqnarray}
where the functions g are flat priors corresponding to a fully uncorrelated parameter set, which range is taken from Ref.~\cite{Margueron2018b}, and $\mathpzc{N}$ the normalization.

The posterior distribution is obtained by  filtering the results of Eq.~(\ref{eq:probalikely})
imposing either physical constraints at high (supra-saturation) density (HD), or ab-initio EFT constraints at low (sub-saturation) density (LD), or both (LD+HD):
\begin{equation}
p_{post}(\{\vec{P_\alpha}\})=p_{lik}(\{\vec{P_\alpha}\})\delta(\mathpzc{F}(\{\vec{P_\alpha}\})-\mathpzc{F}_0),
\label{eq:post}
\end{equation}
where $\mathpzc{F}_0$ is the chosen filter.
The HD filter corresponds to the set of  constraints: (i) positive symmetry energy up to $M_{max}$, (ii) stability of the EoS, (iii) causality up to the maximum mass, (iv) compatibility with the maximum observed  masses $M_{max}\gtrsim 2 M_\odot$~\cite{Antoniadis2013,Arzoumanian2018}, see Ref.~\cite{Margueron2018b} for more details.
The LD filter retains only the EoS passing through the uncertainty band of the MBPT calculations of symmetric and neutron matter by Drischler et al.~\cite{Drischler2016}.
{Other calculations can be found in Refs.~\cite{Tews2016,Holt2017,Tews2018}, which provide comparable theoretical band predictions.}
{ The use 
of a symmetric matter constraint is however important for the determination of the crust thickness, because the transition is   governed by isoscalar instabilities.}

 \begin{figure}[htbp]
     \centering
     \includegraphics[scale=0.43]{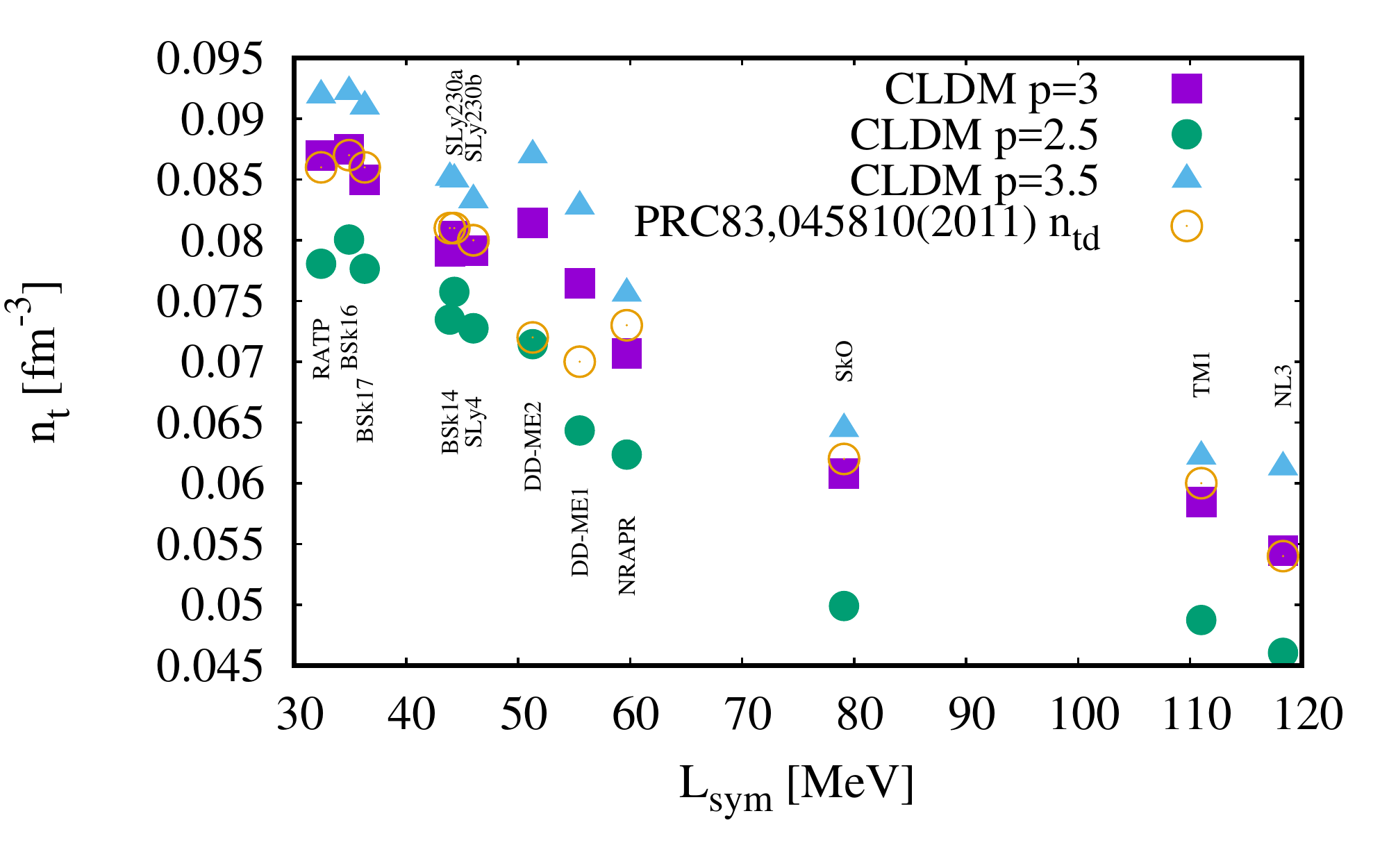}\\
     \vspace{-15pt}
     \includegraphics[scale=0.43]{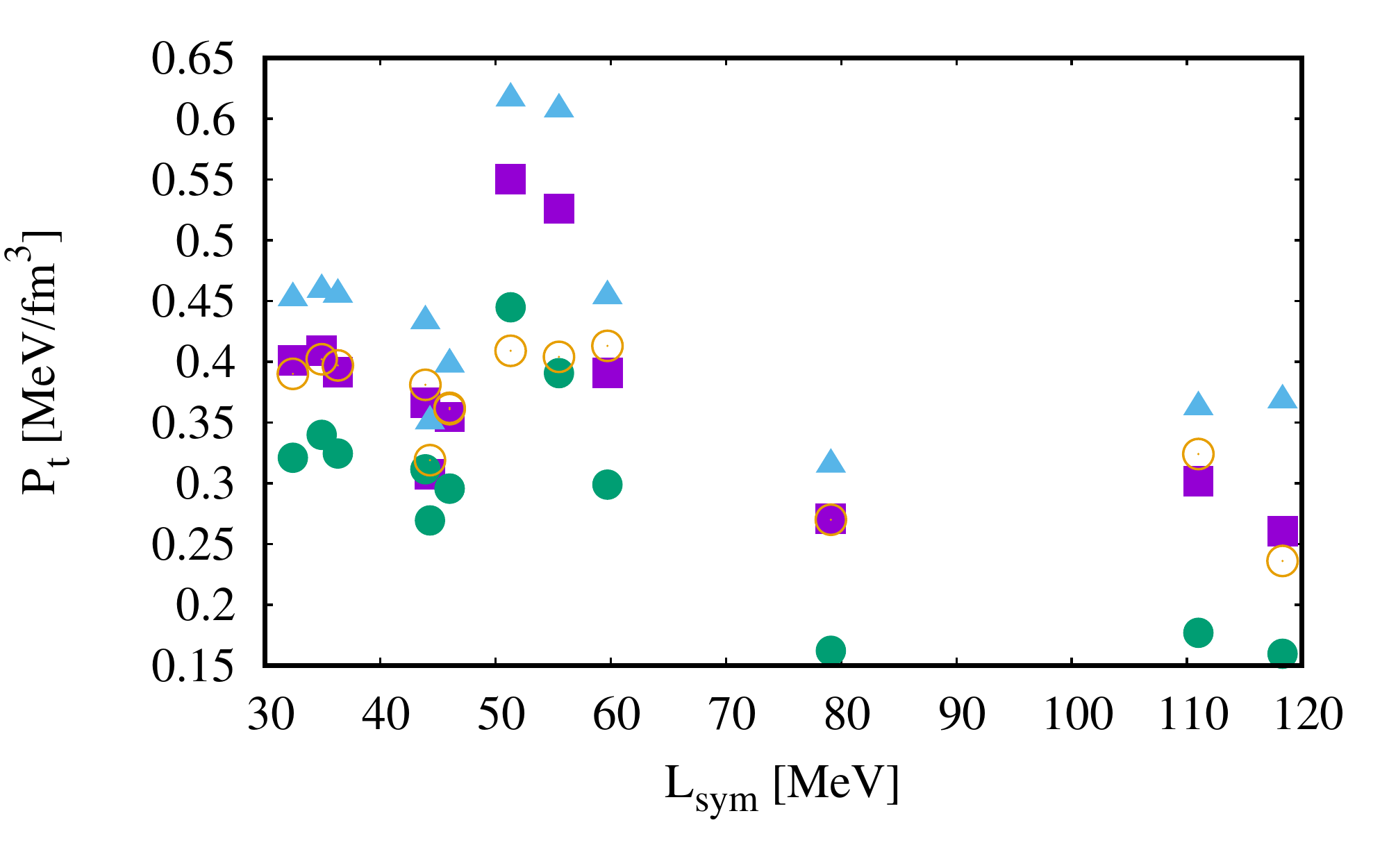}\\
     \vspace{-15pt}
     \caption{(Color online) Transition density $n_t$ (top) and transition pressure $P_t$ (bottom) as a function of $L_{sym}$ for several interactions. The empty dots (squares) are the transition points calculated in Ref.~\cite{Ducoin2011} 
using the dynamical spinodal. The filled circles, squares, and triangles corresponds respectively to our estimation of the transition points with $p=2.5$, $p=3$, and $p=3.5$.
     \label{fig:ducoin}}
 \end{figure}

\begin{figure*}[htbp]
    \centering
    \includegraphics[scale=0.35]{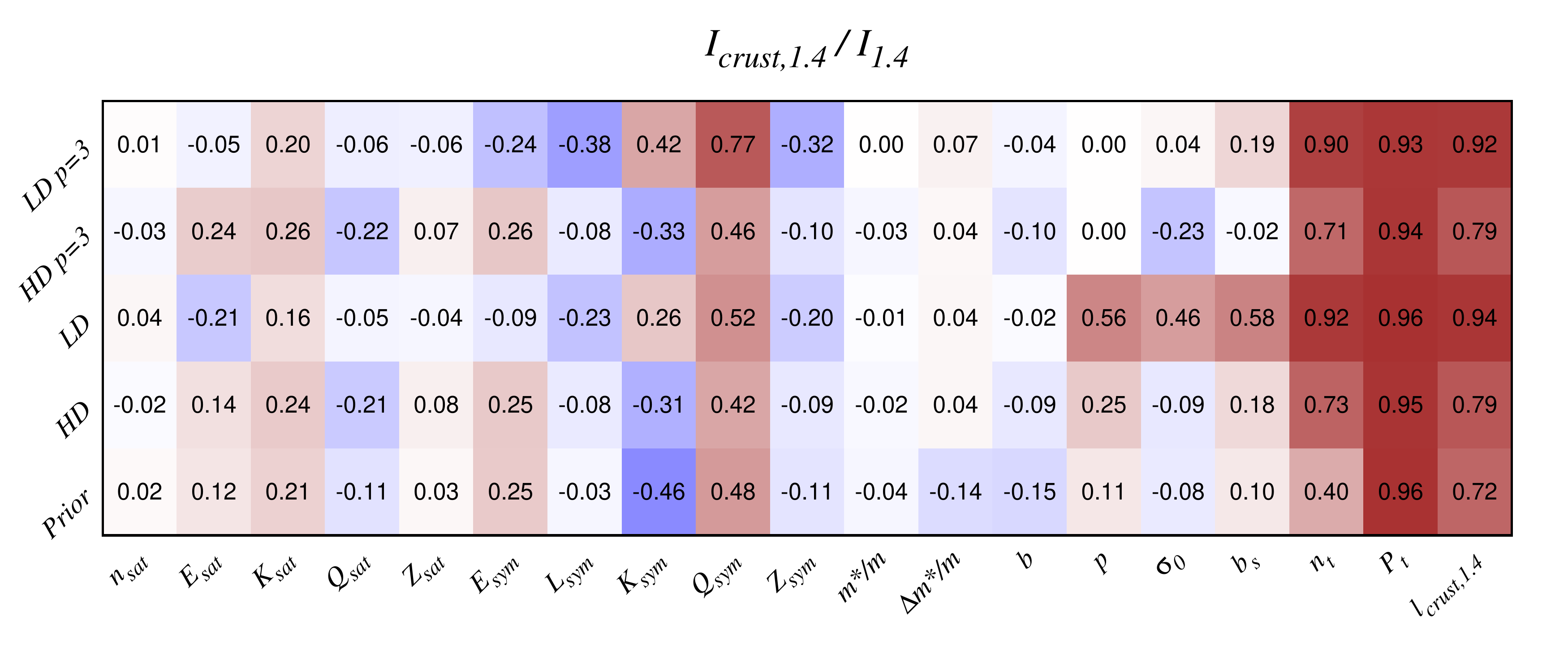}\\
     \vspace{-15pt}
    \caption{(Color online) Correlation between 
the fraction of crust moment of inertia $I_{crust}/I$  for a 1.4$M_\odot$ neutron star and several parameters for different filters. { The red (blue) color scale gives the intensity of the positive (negative) correlation, and the correlation coefficient is explicitly given for each parameter.}
    \label{fig:cor}}
     \vspace{-15pt}
\end{figure*}

Note that the HD filter implicitly implies that first order phase transition { does not occur} 
in the star core up to $2 M_\odot$, as the only hypothesis of the meta-modeling is the analyticity of the EoS \cite{Margueron2018a}. Therefore, imposing the LD filter alone might also be physically acceptable, and we will consider the two filters separately in the following.

Table \ref{tab:ntpt} gives the average values and the standard deviations, defined as
\begin{equation}
    \langle X\rangle = \sum_{\{\vec{P_\alpha}\}} X(\{\vec{P_\alpha}\}) p(\{\vec{P_\alpha}\}),
\end{equation}
for a set of observables $X$.
Fixing $p=3$, we consider different probability distributions 
: the uncorrelated prior distribution $p(\{\vec{P_\alpha}\})=\prod_\alpha g(\vec{P_\alpha})$ (first line), or the posterior distribution Eq.~(\ref{eq:post}) filtered according to the different constraints, $p(\{\vec{P_\alpha}\})=p_{post}(\{\vec{P_\alpha}\})$, see rows 2 to 4.
Knowing the transition point, a numerical solution of the TOV equation allows computing the star radius, the thickness of the crust, and the crustal moment of inertia~\cite{Lattimer2000,Piekarewicz2014}. The first two moments of the  distributions of these quantities are also reported in Table~ \ref{tab:ntpt} for a representative 1.4-solar mass neutron star. 
 The results in Table \ref{tab:ntpt} show that {the high density constraints are essential to establish the average 
crustal properties, but} 
 the knowledge of the low density EoS is very constraining {on the second moment of the distributions.} 
Still, 
 the transition pressure $P_t$ and the fraction of crust moment of inertia have large uncertainties~\cite{Steiner2015} of the order of $34\%$ (resp. $37\%$) considering the LD probability, decreasing to about $28\%$ (resp. $25\%$) if we additionally assume an analytical behavior of the EoS in the full density range covered by the observed neutron star  (LD+HD, see last row in Table \ref{tab:ntpt}).

The isovector surface parameter $p$
plays an important role in the energetics of the inner crust~\cite{Newton2013}, and may depart from its assumed value suggested in Ref.~\cite{Lorentz1993}.
To determine a reasonable prior for $p$, we analyse its effect on the transition point.
Fig.~\ref{fig:ducoin} displays the transition density and pressure obtained 
for a set of relativistic and non-relativistic functionals, 
in comparison with the dynamical spinodal calculation of Ref.~\cite{Ducoin2011}. 
We can see that values of the order $p\approx3$ lead to a general good agreement with the  instability analysis, 
and a variation $\pm0.5$ around $p=3$ provides a good boundary for improved adjustment
\footnote{In the case of the SLy4 functional, the value $p=2.61$ is needed to reproduce the unified EoS approach by Douchin and Haensel~\cite{Douchin2001}.}.  
The impact of varying the isovector surface parameter $p=\{2.5, 3, 3.5\}$ is shown to be quite large in the $1\sigma$ confidence ellipse in Fig.~\ref{fig:sigma_ntpt}.


Which empirical parameters contribute the most to the uncertainty in the observables shown in Fig.~\ref{fig:sigma_ntpt}?
To answer this question, the linear correlation coefficients $r_{XY}=\sigma_{XY}/(\sigma_X\sigma_Y)$ between $I_{crust}$ and the empirical parameters 
$\{\vec{P_\alpha}\}$ are shown in Fig.~\ref{fig:cor}.
{Very similar values for $r_{XY}$ are found for the crustal thickness $l_{crust}$.} 
We can see that isovector empirical parameters are far more influential than the isoscalar ones, as expected, $E_{sym}$, $K_{sym}$ and $Q_{sym}$ being the more influential parameters.
The absence of correlation with the $L_{sym}$ parameter deserves some comments.
It is well known that the NS radius $R$ is well correlated to $L_{sym}$~\cite{Lattimer2000,Steiner2015,Fortin2016}.  The same is true for the core radius $R_{core}$, explaining why the correlations cancel in 
the crustal thickness $l_{crust}=R-R_{core}$ {and consequently on $I_{crust,1.4}$.} 
It then clearly appears that  the higher order parameters beyond $L_{sym}$ must be better constrained to improve the prediction of the crustal properties.

Fixing $p$ tends to increase those correlations, as expected.
However, if $p$ is included in the parameter set, we can see that the uncertainty in  
the surface energy have an impact on the observables shown in Fig.~\ref{fig:cor} comparable to the one of the empirical parameters, see LD row. This is a new feature which has not been reported by previous analyses.

Fig.~\ref{fig:cor} also shows the correlation coefficients between observables.
Large correlations are observed for the transition density and pressure, as expected from previous studies, e.g. Ref.~\cite{Piekarewicz2014}, and the correlation between $l_{crust,1.4}$ and $I_{crust,1.4}$ is also found to be very large, see  Fig.~\ref{fig:sigma_ntpt}.

\begin{figure}[htbp]
    \centering
    \includegraphics[scale=0.531]{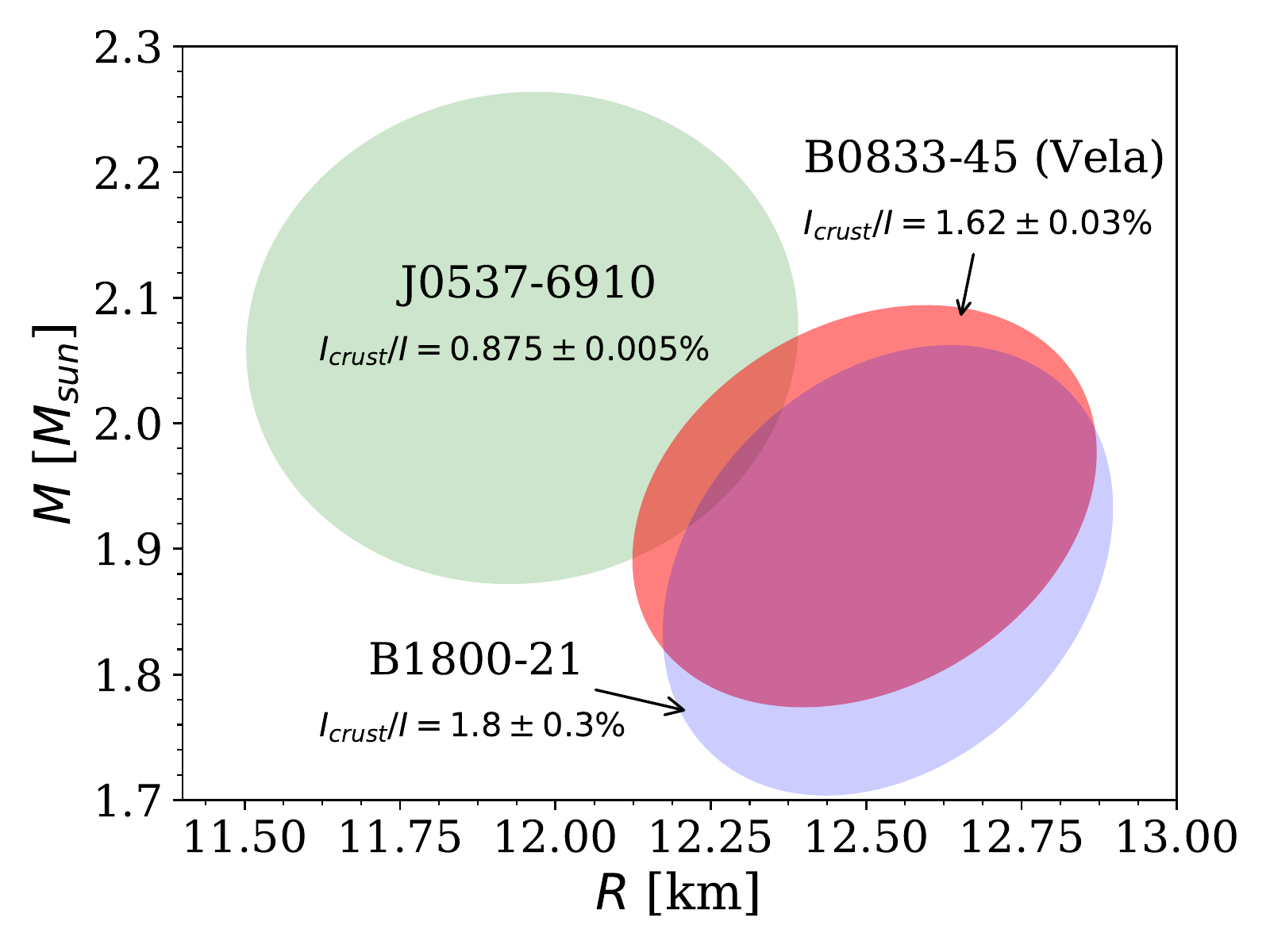}\\
    \includegraphics[scale=0.38]{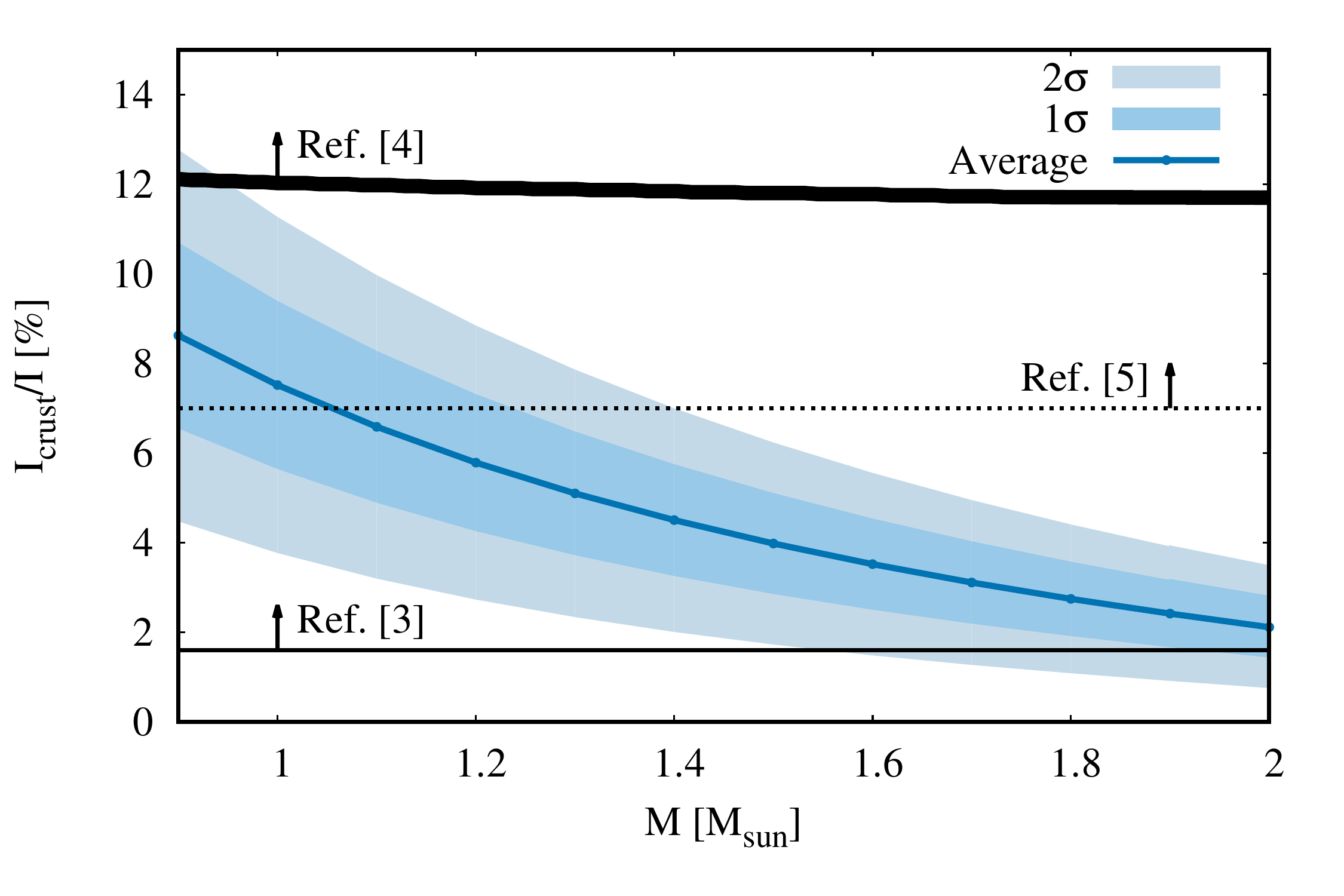}\\
     \vspace{-15pt}
    \caption{(Color online){Top: 1$\sigma$ confidence ellipse with the LD+HD $p=3$ filter for the mass and radius of different pulsars estimated from the observed glitch amplitude from Ref.~\cite{ho}, without crustal entrainment.}
Bottom:
average fraction of crust moment of inertia $I_{crust}/I$ as a function of the mass. The 1$\sigma$ and $2\sigma$ confidence regions are represented, as well as the minimum values needed to justify Vela glitches, with~\cite{Delsate2016,Andersson2012} and without~\cite{Link1999} crustal entrainment.}\label{fig:icrust}
     \vspace{-15pt}
\end{figure}

Finally, we show in Fig.~\ref{fig:icrust} the full impact of our present knowledge on the relation between the {glitch amplitude} 
and the neutron star mass {and radius.  
One-$\sigma$ confidence ellipses 
for three different pulsars estimated from the observed glitch amplitude from Ref.~\cite{ho} are given in the upper part of the figure.
{ Note that an innovative method was proposed to  determine the mass and radius  using observations of the maximum observed glitches~\cite{Pizzochero2017}. It would be interesting to   compare this approach with ours.}
The lower part gives a complete study of the effect of entrainment in the case of Vela:} 
the average value of the fraction of crust moment of inertia $I_{crust}/I$  is shown, as well as the boundaries of the $1\sigma$ and $2\sigma$ probabilities.
The different black lines represent the values proposed with~\cite{Delsate2016,Andersson2012} and without~\cite{Link1999} entrainment effect on the crust moment of inertia to explain  Vela glitches.
From Fig.~\ref{fig:icrust} we can conclude that the value determined for the maximal entrainment effect is incompatible with the present nuclear physics knowledge.


In conclusion, considering the experimental and EFT theoretical predictions at low density,
the uncertainty on the crust thickness (relative moment of inertia) is of the order of 9\% (25\%), for $M=1.4 M_\odot$.
These uncertainties originate from the dispersion in the predictions of the crust-core transition point, which in turn depends on the high order isovector empirical parameters 
$K_{sym}$ and $Q_{sym}$, as well as on the isovector surface energy parameter $p$.   
Higher precision in the experimental determination of $K_{sym}$ and $Q_{sym}$, in the low density EFT theoretical predictions, and in the microscopic modeling of the surface energy at extreme isospin ratios are needed to reduce the uncertainties of crustal observables.

\begin{acknowledgments}
This work was partially supported by the IN2P3 Master Project MAC, "NewCompStar" COST Action MP1304, PHAROS COST Action MP16214.
\end{acknowledgments}


\end{document}